\documentclass{article}
\usepackage{hyperref}
\usepackage{amscd,amsmath,amsthm,amssymb}
\usepackage{enumerate,varioref}
\usepackage{epsfig}
\usepackage{graphicx}
\usepackage{subfig}
\usepackage{mathtools}
\usepackage{tikz}
\usepackage{algorithmicx}
\usepackage{algorithm}
\usepackage{algpseudocode}
\usepackage{xcolor}

\theoremstyle{definition}

\theoremstyle{remark}

\title{Portfolio Optimization of 60 Stocks Using Classical and Quantum Algorithms}
\author{Chicago Quantum\footnote{Jeffrey Cohen, Alex Khan, Clark Alexander} \\ email \href{mailto:research@quantum-usaci.com}{the authors}}

\begin{document}

\maketitle

\tableofcontents

\begin{abstract}
We continue to investigate the use of quantum computers for building an optimal portfolio out of a universe of 60 U.S. listed, liquid equities. Starting from historical market data, we apply our unique problem formulation on the D-Wave Systems Inc. D-Wave 2000Q\textsuperscript{TM} quantum annealing system (hereafter called D-Wave) to find the optimal risk vs return portfolio.  We approach this first classically, then using the D-Wave, to select efficient buy and hold portfolios. Our results show that practitioners can use either classical or quantum annealing methods to select attractive portfolios. This builds upon our prior work \href{https://arxiv.org/pdf/2007.01430.pdf}{on optimization of 40 stocks}
\end{abstract}

\section{Introduction}
Our work is inspired by the notion that we can find attractive investment portfolios from a universe of US equities leveraging a quantum computer.  As we scale the problem with the number of equities analyzed (portfolio choices are $2^n$), we investigate whether quantum annealing can scale up vs. classical methods and select a reasonable sized grouping of attractive portfolios, as opposed to just one ideal solution.

We advanced our classical methods.  We now have five methods to find the ideal portfolio from 60 stocks in under a minute.  This raises the bar for quantum annealing, which takes about 0.13 seconds (or $\sim 130,000 \mu s$) to sample 200 to 500 times within one run (to find the best $N$ out of 60 stocks).  The D-Wave time, 21 seconds, is an accumulation of all runs used in this research in our primary account, and the results are an accumulation of all valid portfolios found in this research.  

We retain our prior formulations; The Chicago Quantum Net Score and the Chicago Quantum Ratio defined by the following

\begin{equation}
\text{CQNS}(R_w, \alpha) = Var(R_w) - \mathbb{E}[R_w]^{2+\alpha}
\end{equation}

and 
\begin{equation}
\text{CQR}(R_w) = \frac{Cov_{im}(R_w)}{\sigma(R_w)}
\end{equation}

where $R_w$ is a weighted portfolio, $\alpha > 0$ is a real number, which we generally set to 1.  $Cov_{im}$ is the covariance of our portfolio against the entire market, which we take as the S\&P 500 for this article.

In this paper we will provide our progress, and setbacks, with classical methods, then focus attention on our quantum annealing on the D-Wave DW\_2000Q\_6 2,048 qubit, D-Wave 2000Q lower-noise system, with the [16,16,8] Chimera topology. We will then provide a comparative analysis and discuss the scale-up potential of the D-Wave. 

\section{Market Context}
We performed our research during a time of market increases for the largest companies, and a relatively low interest rate environment.  Our analysis used a risk-free rate of 1\%.  The markets have seen a rise of 18.60\% over the past year, when taken as the average of the increases of the four equity indices we use: S\&P 500, Nasdaq Composite, Russell 2000 and Wilshire 5000.
The range of $\beta$ for the 60 stocks was $[0.417, 2.12]$.  The variance of the S\&P 500, used in the Chicago Quantum Ratio, is 0.00045105.  The 60-asset, all in portfolio of equally weighted stocks, has an expected future return of 22.09\% (using the Capital Asset Pricing Model), and the standard deviation is expected to be 2.48\%, yielding a Sharpe Ratio of 8.92.  Our model does use prior year trading history to pick its portfolios. The parameters of our model remain unchanged since our last paper in July 2020.

\section{Shape of the Energy Landscape}

\begin{figure}[!ht]
	\centering
	\includegraphics[width=0.9\textwidth]{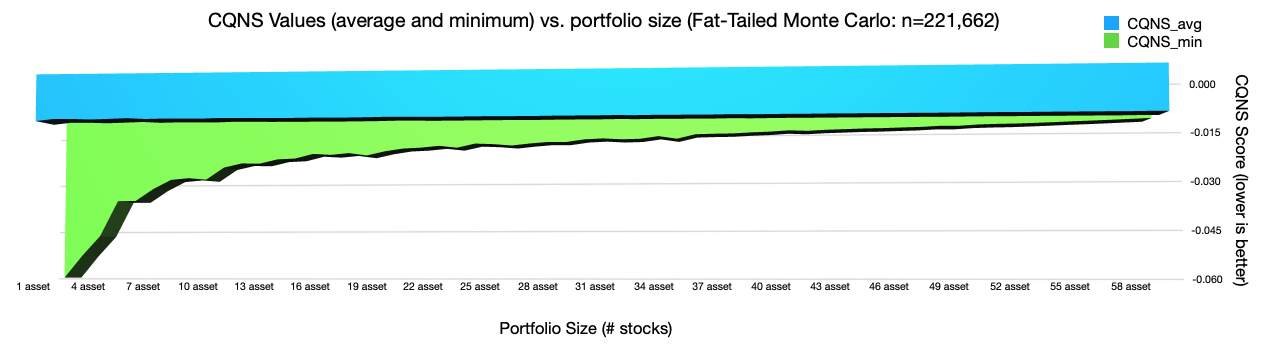}
	\caption{CQNS minima and averages}
	\label{fig: 1}
\end{figure}

We start with a Monte Carlo random sampling analysis to understand the high-level shape of the energy landscape before we look to the different methods to solve it.  We score in excess of 220 thousand portfolios across all portfolio sizes and store the average, minimum and maximum CQNS values which we can plot and review.  As shown in \ref{fig: 1} the blue bar shows the CQNS average, which is stable across portfolio sizes, which implies there is no bias toward any portfolio of a particular magnitude.  Additionally, in green we see the CQNS minimum values which have a few special cases, which are highly attractive portfolios where the expected return greatly outweighs the expected risk.  The shape may be different for each set of stocks analyzed.

\section{Classical and Hybrid Methods}
\subsection{Fat Tailed Monte Carlo Analysis}

We keep track of our random samples in an array which holds the minimum, maximum and average CQNS score for each size portfolio analyzed.  We initialize this with random samples in a discrete distribution (centered around $N/2$ assets), then run a random sample for each portfolio size.  In our research, these two sampling methods generated 221,660 samples.  In our last run, we found the ``ideal" portfolio as the solution is in a very small portfolio.

This ``fat tailed MC" method will do well if the solution is either very large or very small.  The random sampler which is centered on the $N/2$ size portfolios, will generally not do well.  In a 60 asset universe, there are $1.15 \times 10^{18}$ possible combinations.  We cannot count on sampling all of them.  However, we ran this experiment twice.  In one case we found the best answer and in another we found the $2^{nd}$ best answer; both in 24 seconds.  Again, this only worked because the best answer was in a tail of the portfolio sizes with very few possibilities, and the random sampler can find them.

Like in the simulated annealing case, the Monte Carlo analysis has the greatest range of CQNS values at the smallest portfolios (best and worst portfolios).  The range and relative attractiveness narrows as we analyze larger portfolio sizes.  In the graphics, the sequence moves from small to large portfolios.

\subsection{Genetic Algorithm}

A genetic algorithm looks to bring the best attributes (stocks) forward from two portfolios (parents) by a process of breeding them and creating new portfolios (children and mutations) that we then score.  We keep the best X portfolios as scored by the CQNS as parents in the next generation, then breed again.  

Our genetic algorithm (GA) is custom coded to start with an initial population of parents (either 456 random portfolios or seeded with D-Wave solutions).  We tune it to run for 40 generations and pass the 40 best solutions (equal or better values only) to the next generation.  We breed a ratio of 3:2 children to mutations.  Experimentally, both solve the problem of optimizing the 60 asset portfolio quickly.  In our last run, the GA (456 random) took 7 seconds and GA (2,588 D-Wave) took 48 seconds to find the best solution.  Typically, however, the GA (D-Wave) runs 20\% faster than GA (random) as we start with a smaller and better scoring initial population.

\subsection{Simulated Annealer}

A simulated annealer models the temperature-based evolution process, where the algorithm is looking for the lowest energy solution where the ability to jump to a new interim solution outside of a local minima decreases as the system cools.  At higher temperatures, it is more free to tunnel or jump to neighboring energy levels and look for deeper energy minima.  As the temperature ``cools", it becomes harder to jump far and we look for the best solution in that neighborhood.  In this model, the Chicago Quantum Net Score is a fixed multiple of the energy level at each ($N$ of 60-asset) QUBO.  By minimizing the energy level, we find the best CQNS and investment portfolio.

Our simulated annealer is custom coded in Python 3 in about a page of code.  In our tuned version, it finds the optimal 60 asset solution some of the times, and a good solution otherwise.  We can tune it to run longer, which increases the frequency of success.  In our last run, it found the optimal portfolio from 60 stocks in 15 seconds.  

We tune our simulated annealer with four parameters.  (1) When to jump to a neighboring solution.  When the temperature is warm, sometimes our algorithm jumps even when the score is slightly worse.  (2) Initial temperature and minimum temperature which determines the temperature range for annealing.  (3) Cooling rate determines how fast the temperature cools per cycle, and is used to determine the number of annealing cycles $\frac{T_{max}- T_{min}}{\text{cooling rate}}$, 
and (4) the number of annealing trials, or as D-Wave calls them ``sweeps" per annealing cycle.  We tune this mix to maximize the frequency of finding the optimal solution in the shortest time.  This is an ad-hoc exercise. 

We also use the D-Wave Simulated Annealer as an alternative sampler.  In our most recent runs, it finds the optimal portfolio that we allowed it to find in 11 seconds, but modifies the energy level of that portfolio.  We run the simulated annealer by specifying the range of portfolio sizes, and resulting QUBO, to run through the annealer.  Normally we run $[2,59]$, but in the last case to save time we ran $[2,50]$.  
Like the quantum annealer, it has to match the number of assets to the desired portfolio size for us to accumulate those answers.  

Our settings have been tuned to speed up the annealer.  We run a $\beta$\_range, or inverse of the temperature, at $[0.000001, 9]$.  We run the simulated annealer to sample 200 portfolios for each portfolio size.  We also set the number of ``sweeps" or times to look at each energy cycle at 200.  We use a $\beta$ schedule of type ``geometric."  With these settings, we run for 11 seconds, find 168 out of 9,600 trials, and see the $2^{nd}$ best result (and the best we allow it to find), repeatedly in the results.  If we want to look at good portfolios at different sizes, like we can with the quantum annealing answers, we see valid portfolios at many portfolio sizes.

In the graphic below we see that the smaller portfolios (run in order of size), have the widest spread, and the best possible results. As we get larger, we see an almost asymptotic tightening of the CQNS scores in the middle.  We see this with most samplers we use.

\begin{figure}[!ht]
	\centering
	\subfloat[Simulated Annealing Solutions]{{\includegraphics[width=4.9cm]{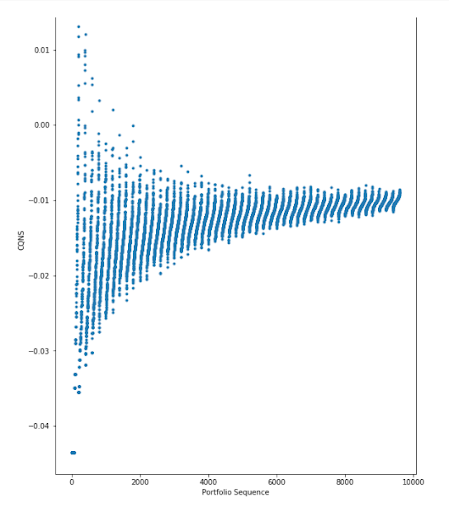} }}%
	\qquad
	\subfloat[Monte Carlo Simulation]{{\includegraphics[width=5.5cm]{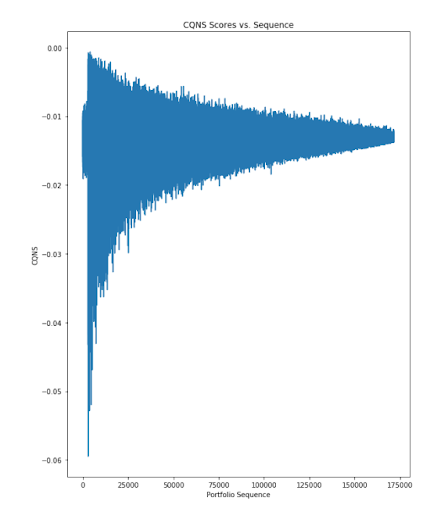} }}%
	\caption{Simulated Annealing vs Monte Carlo Sampling}%
	\label{fig:SimAnnealMC}%
\end{figure}

\subsection{D-Wave Tabu Multistart MST2 Sampler}

The D-Wave Tabu Sampler was run against our QUBOs and we saw the least attractive portfolios from this method.  The sampler picked most of its solutions from the 20 to 40 asset size portfolios despite being run against the different QUBOs with the penalty functions.  We also see that the Tabu sampler started by finding very large (poor) CQNS values, and quickly reached a plateau of its best answer.  Unfortunately, the best Tabu scores found were worse by a factor of 10 or more.  

\begin{figure}[!ht]
	\centering
	\subfloat[Assets Chosen vs QUBO size]{{\includegraphics[width=4.9cm]{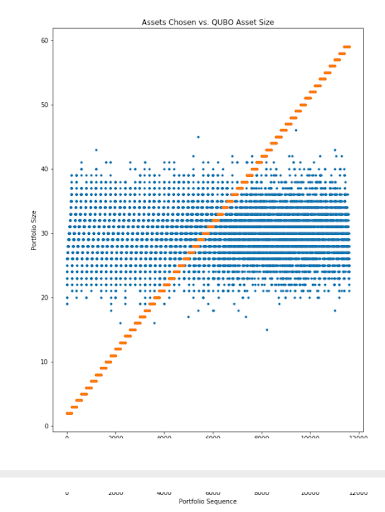} }}%
	\qquad
	\subfloat[Portfolio Sequence]{{\includegraphics[width=5.5cm]{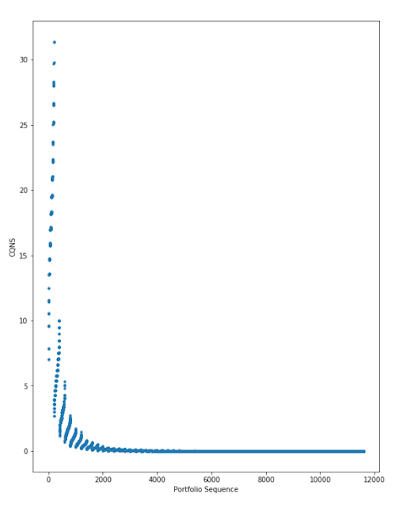} }}%
	\caption{On the right, we see the Tabu sampler picked portfolio sizes between 20 and 40 regardless of the desired portfolio sizes, and penalty functions.  On the left we see CQNS values chosen by the Tabu sampler, and how it starts with poor (high) values.}
	\label{fig:AssetsChosen}%
\end{figure}

We set the sampler to run from $[2,60]$ assets, with 200 reads per QUBO, a scale\_factor of 1, a maximum of $20 \mu s$ to run, and a tenure of 50.  The tenure is the number of answers to store in memory to save time in the run.  Like with the D-Wave quantum annealer, we keep and accumulate valid portfolios from each size portfolio, and in 11,600 trials we had 190 valid portfolios found.  The final run took 267 seconds.  We have an open question of whether increasing the run time, reads per QUBO, and lowering the tenure would help us find better answers.  We had initially set the Tabu Sampler to run longer with these settings, but did not see better results.  

\subsection{D-Wave Hybrid Sampler}

At this point we do not see valid results from the hybrid sampler.  It is more of a ``black box" solver for us where we can set a few parameters and feed it the same $60\times60\times60$ matrix used by the other methods.  We set it to run across portfolio sizes using the same QUBO as the other methods, and it found no valid portfolios that match the size required.  It does find ``good"’ portfolios, but the CQNS scores are incorrect due to the penalty we apply.  

\section{D-Wave Systems Quantum Annealer}

\begin{figure}[!ht]
	\centering
	\subfloat[10 assets]{{\includegraphics[width=5.5cm]{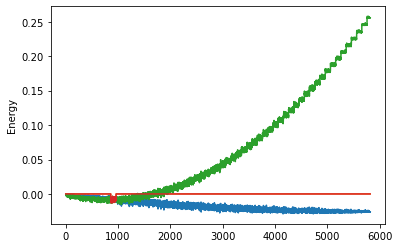} }}%
	\qquad
	\subfloat[20 assets]{{\includegraphics[width=5.5cm]{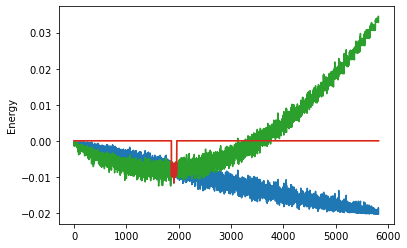} }}%
	\newline
	\subfloat[30 assets]{{\includegraphics[width=5.5cm]{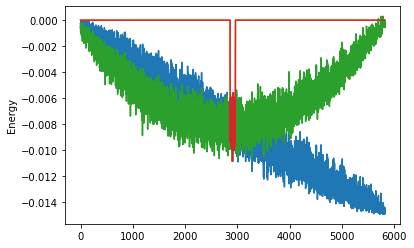} }}%
	\qquad
	\subfloat[40 assets]{{\includegraphics[width=5.5cm]{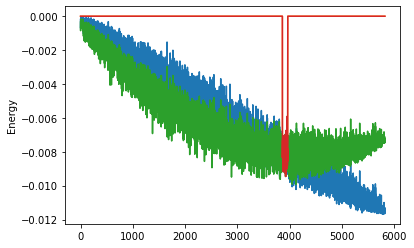} }}%
	\newline
	\subfloat[50 assets]{{\includegraphics[width=5.5cm]{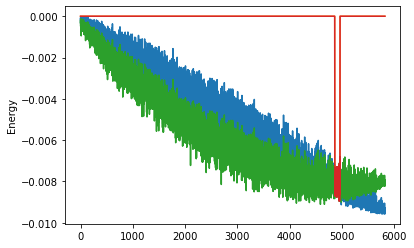} }}%
	\qquad
	\caption{Charts showing resulting energy levels from QUBO ($N=10, 20, 30, 40, 50$ assets) before (blue) and after (green) affine transformation based on random sampling.
	}
	\label{fig: Shifting to capture assets}%
\end{figure}

Similar to our simulated annealer, a quantum annealer modifies the quantum fluctuations of a physical set of qubits to allow the spin of the qubits to ‘jump’ to a neighboring solution to find a lower energy level.  The resulting energy level, which we convert to a CQNS score, is the value the quantum annealer is minimizing by making those jumps between solutions.  At the end, the system reads the energy and spin of the qubits, rounds up or down, does a majority vote if there is a chain break, and outputs a string in $\{0,1\}^{60}$  (one for each stock), the energy of the system, and the value of the chain break, if any.  We re-factor the energy to a CQNS score, and append the values to an output file.

We build a $60\times 60\times 60$ ``bigmatrix" of values which load the negative of the expected return content on the linear terms, and load the variance and covariance content on the quadratic terms. 

We then apply an affine transformation to the values (shift) so the non-desired number of assets have a penalty.  We set the affine transformation with three points.  The zero asset point must have an energy value of zero, the desired asset count must have an accurate count, and we can increase the energy level for all other points.  We create a shape like a pulley in the energy values that result from executing the QUBO in simulation or in D-Wave.

In figure \ref{fig: Shifting to capture assets} we see the 10, 20, 30, 40 and 50 asset QUBO energy landscape.  We bend the shape of the curve so the desired number of assets is centered around the lowest energy levels, while still staying accurate at that number of assets, during a small random sampling of CQNS values.  We ground the 0,0 point for zero assets, set the point for the number of assets desired to be accurate, and vary the energy level of the 60,60 asset point to the positive, or a penalty.

Sometimes creating an affine transformation with the apparent shape we need is not enough.  In other words we might not get any matching results for the desired assets. As we run our QUBO on D-Wave, we also visualize the results of all the samples received back  from D-Wave.  We use that visualization to further tune the shift of the matrix.  In the chart below, we see the orange points which indicate the assets desired by each QUBO run.  The blue points are the size of the portfolios returned by D-Wave.  In case of overlap, we see the orange dots.  You can see that as we exceed 36 assets, the portfolios received were significantly smaller than desired, which required us to further adjust our shift values.

\begin{figure}[!ht]
	\centering
	\subfloat[Quadratic Affine Transformation of trials size 2 to 50]{{\includegraphics[width=5.5cm]{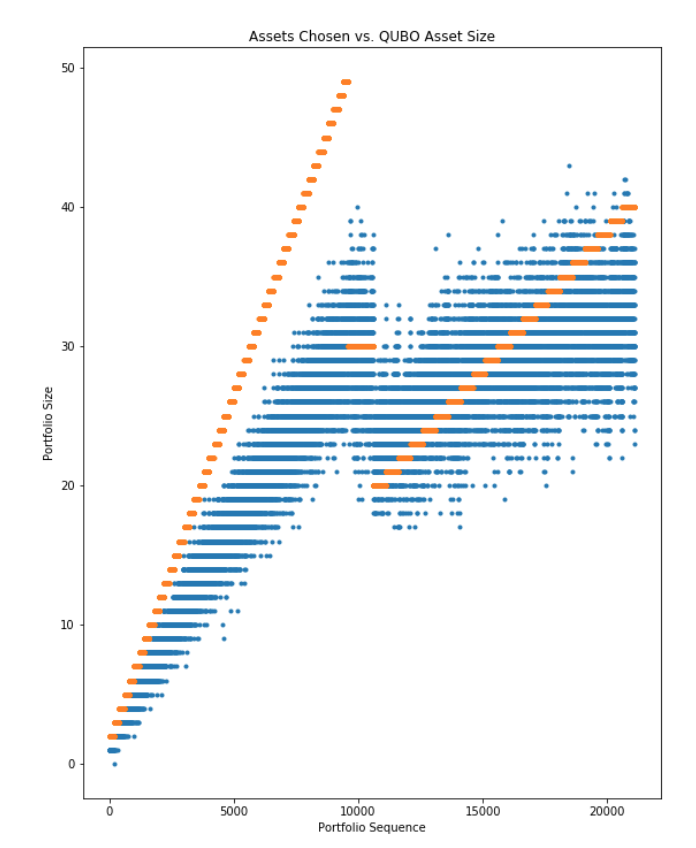} }}%
	\qquad
	\subfloat[Graduated Affine Transformations of trials size 2 to 50]{{\includegraphics[width=5.0cm]{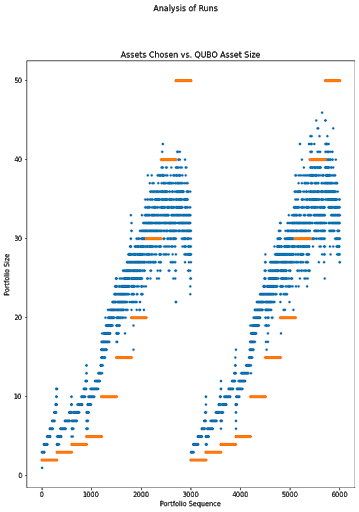} }}%
	\caption{On the left: two trials: 2-50: resulting portfolios always lower than desired number of assets. 20-40: after shift shows results in the number of assets desired. On the right: displaying how a small shift in the affine transformation allowed portfolios detected at the higher end 40 assets to increase. The desired effect is having as little spread as possible around a centered orange line.  
	}
	\label{fig: Affine}%
\end{figure}

After we apply the affine transformation, we scale the matrix values to between $[-0.9, 0.9]$ before we send it to D-Wave.  Experimentally we find D-Wave does better when we apply the initial scaling to the data.  We believe this is because D-Wave does additional scaling while it feeds the data to the quantum annealer, and if the values are divergent enough may apply different scaling factors to the linear and quadratic terms.  This would nullify our CQNS score and the predictive ability of our model.  The D-Wave annealer can accept H\_Range values between $[-2,2]$ in the linear term and J\_Range values between $[-0.9,0.9]$ in the quadratic terms for Ising formulations.  We find that if we perform our scaling, we receive valid D-Wave energy levels, and resulting CQNS scores when compared to classical validation.

We find it difficult to have D-Wave find larger-sized portfolios.  It typically finds smaller portfolios despite the penalty on them.  In our research on 60 assets, the largest valid portfolio we received from the D-Wave was 47 assets.  We believe we could continue to adjust chain strength and annealing time to increase our chances of a matching portfolio. However, we believe we may need to further refine the QUBO for larger portfolio sizes (e.g., 45 or more assets) to continue to scale to 100 or more assets.  This assumes we believe the largest portfolios can also be the most efficient. 

The challenge is our quadratic values become too small relative to the linear terms.   At 59 assets, the linear terms still vary from $[-1,0.2]$, but the quadratic terms are $[0.0037, 0.0045]$.  

\begin{figure}[!ht]
	\centering
	\subfloat[bigmatrix for 59 assets]{{\includegraphics[width=10cm]{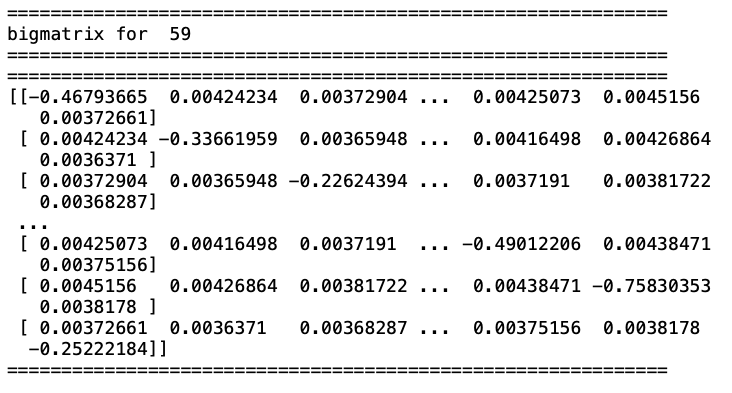} }}%
	\qquad
	\newline
	\subfloat[heat map for big matrix with 59 assets ]{{\includegraphics[width=5.5cm]{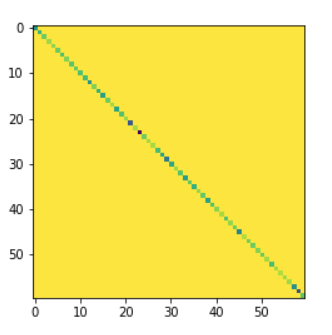} }}%
	\caption{Heat map for a big matrix with 59 assets. 
	}
    \label{fig: bigmatrix59}
\end{figure}
We understand why this happens.  The covariance is a quadratic term and it decreases as the number of assets increases. At just a few assets, the linear and quadratic terms are in the same order of magnitude, and D-Wave seems to find excellent, or in one case the ideal value, in those smaller assets.  

\begin{figure}[!ht]
	\centering
	\subfloat[bigmatrix for 2 assets]{{\includegraphics[width=10cm]{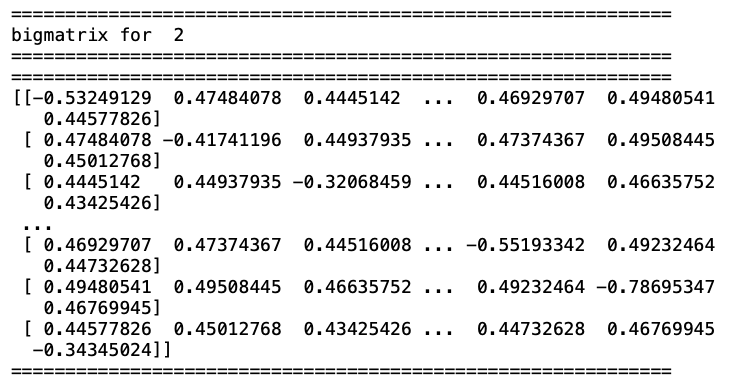} }}%
	\qquad
	\newline
	\subfloat[heat map for big matrix ]{{\includegraphics[width=5.5cm]{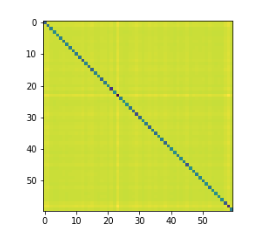} }}%
	\caption{Heat map for a big matrix with 2 assets. 
	}
	\label{fig: bigmatrix2}%
\end{figure}

This is an open question in our continuing research...how to rescale or reformulate our QUBO so that portfolio sizes from $N/2$ to $N$ have increasing covariance values on the quadratic terms instead of decreasing as they do now.  

We then convert the $60\times60\times60$ shifted matrix to QUBO formulations which are individually sent to D-Wave. D-Wave applies its own scaling to achieve acceptable H\_range and J\_range values.  This same matrix is used for all of our classical models which require a matrix as input (e.g., simulated annealer, TABU sampler, and the hybrid solver).

We run the quantum annealer repeatedly against our QUBOs in this experiment and accumulate valid portfolios (where number of assets chosen matches the number of assets desired in that QUBO).  Our initial runs, which accumulated 116 values, were on the 30 asset QUBO and were intended to calibrate and tune the D-Wave and parameters we use.  We adjust settings for chain strength, post processing, shift, scale, annealing time, reduce intersample correlation, and to look at the resulting solutions, and information about the programming of the QUBO on the qubits.  We use the D-Wave Inspector to better understand the solution space and the loading of the QUBO.  Including the calibration, we accumulate a total of 3725 valid portfolios with portfolio size ranges of $[2,47]$.

\begin{figure}[!ht]
	\centering
	\includegraphics[width=0.9\textwidth]{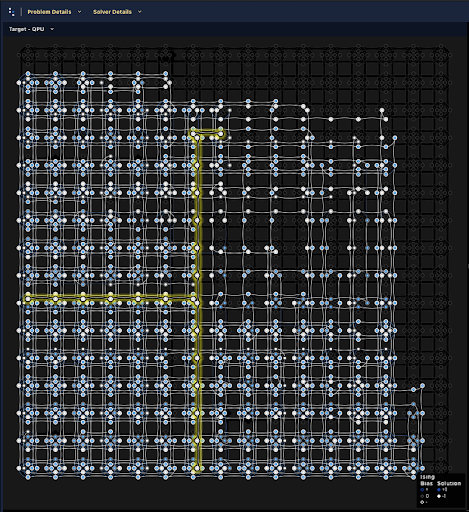}
	\caption{The connectivity of D-Wave in the [16,16,8] topology for 58 of 60 assets}
	\label{fig: dwave topology}
\end{figure}

We find an inverse relationship between the chain strength setting and the maximum chain length (and resulting chain breaks).  We find an inverse relationship between chain strength and sensitivity to find valid portfolios.  We settled into a chain strength range of $[0.2, 2.0]$ against a default value of 1.0.  Chain strengths $> 2.0$ did not provide valid results in our runs.
In this chart you can see the layout of our $60\times 60$ QUBO shifted for 58 assets after embedding \ref{fig: dwave topology}, or programming, the QUBO onto the D-Wave qubits and before the annealing process begins.  The yellow line traces the path of one asset being embedded on multiple qubits in a chain, with covariance connectivity to all other stocks. This embedding is required to ensure our covariance relationships are retained with the limited number of connections available between qubits in the actual hardware.

We also found a relationship between annealing time and number of valid portfolios found in each run.  We vary our annealing time in $[20, 200] \mu s$, against a default of $20 \mu s$.  The best results came from larger portfolios $[40, 47]$ and the smallest portfolios $[2,4]$ at $100\mu s$ and chain strength of 2.  We did not try annealing times $>200 \mu s$, although they can go as high as $2,000 \mu s$.  Longer annealing times, and higher chain strength values, might be the way to produce valid solutions at the largest portfolios $[48,59]$.  The mid-sized portfolios $[7-24]$ did best with a chain strength of 0.2 and a fast annealing time of $20 \mu s$.

\section{Quality of Solutions Found}

\begin{figure}[!ht]
	\centering
	\includegraphics[width=0.9\textwidth]{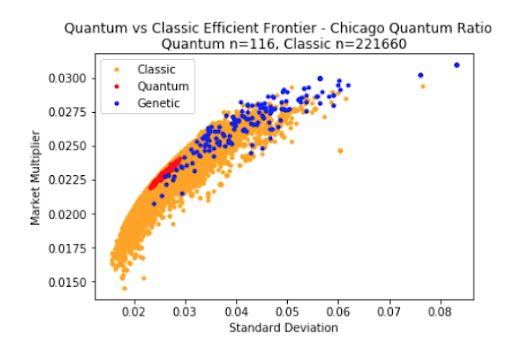}
	\caption{Several Methods using CQR}
	\label{fig: CQRvGenetic}
\end{figure}

\begin{figure}[!ht]
	\centering
	\subfloat[CQR]{{\includegraphics[width=5.5cm]{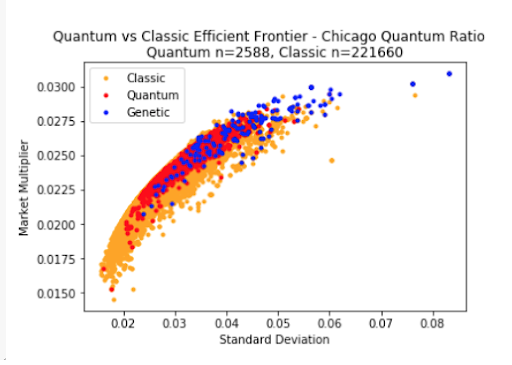} }}%
	\qquad
	\subfloat[CQNS]{{\includegraphics[width=5.5cm]{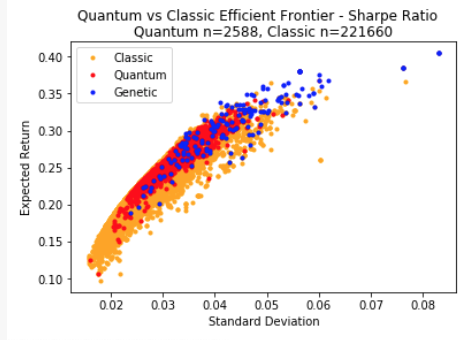} }}%
	\caption{Several Comparative methods with different scores.}
	\label{fig: CQRvCQNS}%
\end{figure}

In this paper, we use the CQNS, and the related Chicago Quantum Ratio, as a proxy for the classical method of portfolio optimization.  We consistently find that the D-Wave system, with just a few valid portfolios, picks portfolios that are on or ahead of the efficient frontier as found through random sampling.  In this case, 414 valid D-Wave quantum annealing portfolios compare favorably against 40,000 random portfolios, and outperform at the higher risk (or standard deviation) levels seen below in \ref{fig: CQRvGenetic}\ref{fig: CQRvCQNS}.  
  When we build investment portfolios, we find the D-Wave selected CQNS portfolios tend to be relatively more risky than CQR portfolios while both are still highly efficient.

\begin{figure}[!ht]
	\centering
	\includegraphics[width=0.75\textwidth]{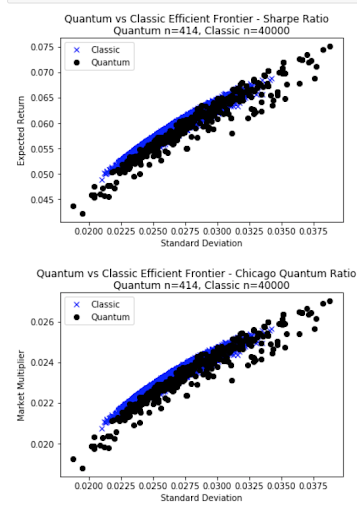}
	\caption{Above: The comparison of 414 quantum portfolios vs 40000 Monte Carlo samples using the Sharpe ratio. Below: The comparison of 414 quantum portfolios vs 40000 Monte Carlo samples using CQR}
	\label{fig: 9}
\end{figure}

In the graphic\ref{fig: CQRvGenetic}, you can see the initial 116 valid portfolios with 30 assets (out of 60) that we used for calibration compared to our 221,660 Monte Carlo results.  The quantum portfolios, in red, have relatively low CQNS values (-0.0146 vs. -0.0437 for top 10 portfolios). D-Wave appears to be picking efficient portfolios, even out of a population of average results. The blue dots reflect the genetic algorithm and the simulated annealing results.  They reflect the best, or ideal values, and a sample of the values from the previous generations.

After accumulating 2,588 valid portfolios, we can see that the red dots cover a significant portion of the efficient frontier.  The quantum points provide investors with a broad coverage of the efficient frontier after selecting just 2,588 portfolios.

\section{Comparative Analysis}

\begin{figure}[!ht]
	\centering
	\includegraphics[width=0.95\textwidth]{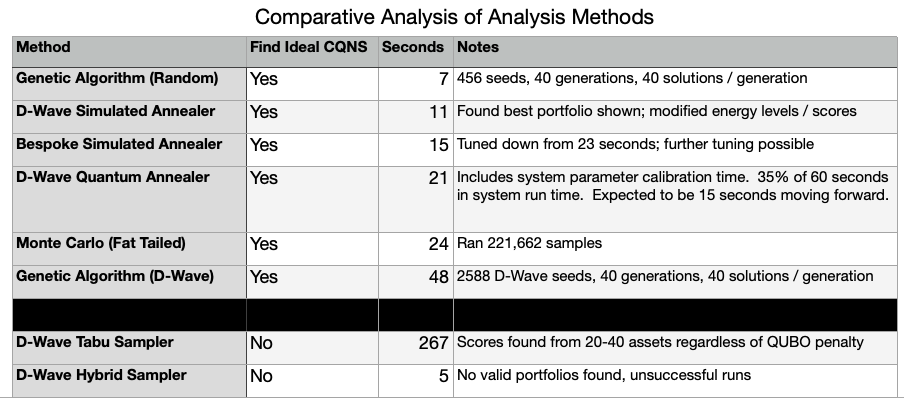}
	\caption{Table of Comparative methods}
	\label{fig: comparison table}
\end{figure}

\begin{figure}[!ht]
	\centering
	\includegraphics[width=0.95\textwidth]{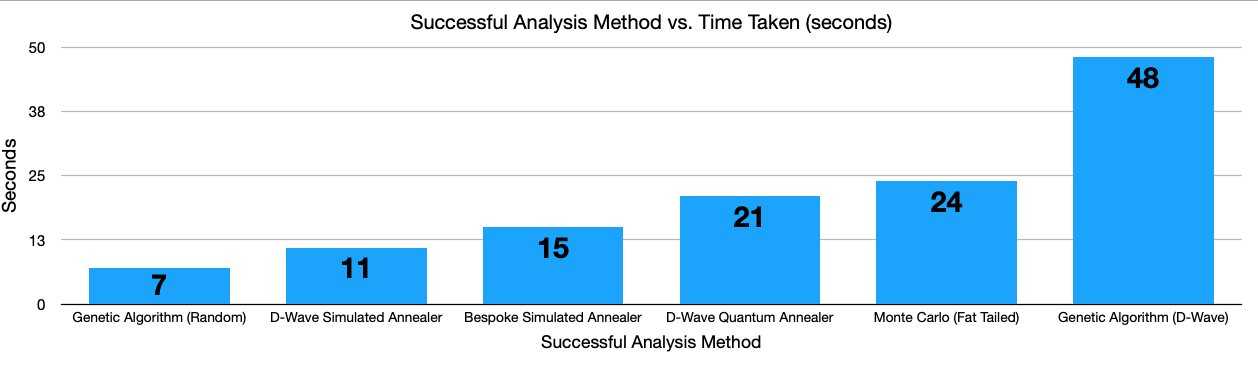}
	\caption{Analysis of Comparative methods}
	\label{fig: method table}
\end{figure}

We measure success in whether the different methods find the ideal portfolio and how long each method takes.  Six methods \ref{fig: comparison table}, including D-Wave, found the ideal portfolio, which is the portfolio we believe has the lowest (best) CQNS score.  For four of the methods \ref{fig: method table}, we had Python time the run duration.  For the D-Wave quantum annealer we used the percentage of account resources used (against a 60 second budget) for the primary account used. The classical methods were run on a 2013 iMAC with a 3.4GHz Quad-Core Intel Core i5, with 16GB DDR3 RAM at 1600 MHz, running macOS Catalina and Python 3.7.7. 

\section{Scale-Up Potential}

We see a few challenges in scaling up this analysis from 60 stocks to 100, and up to 500 stocks to be analyzed at once.  
First, our formulation puts significant strain on the system to evaluate portfolios with larger numbers of stocks (47+).  The variance and covariance content between each stock (and loaded onto qubit coupling terms) becomes too small to be effectively read by the D-Wave system.  Values currently approaching 0.003 as compared to linear terms of -0.9. Second, room on the D-Wave QPU is filling up.  When we program and embed 60 stocks, we use between 1,300 and 1,700 of the available 2048 qubits.  Using a linear extrapolation, there may only be room for 20\% more, or 72 stocks, at the higher end of the loading.


\section{Conclusion}

Our aim is to extend our research from a 40 to 60 asset universe, to find the ideal portfolio out of a possible $2^{60}$ options.  Using the CQNS formulation we extend to 60 assets on both the D-Wave quantum annealer and five classical methods.  We do this through engineering improvements and applying a graduated penalty to the $60\times 60\times 60$ matrix.  We improve our classical methods and all run in under one minute.  

We believe we can scale the problem further with improved embedding, management of the energy delta between portfolios, and continued improvement in the penalty function.  

Our results show that D-Wave can be leveraged to generate an efficient frontier landscape quickly with a small number of samples (e.g., $<1,000$) and create a set of attractive portfolios, while potentially finding the ``ideal" portfolio.  Our classical methods continue to validate the quantum results.

\section*{Addendum}

\begin{figure}[!ht]
	\centering
	\includegraphics[width=1\textwidth]{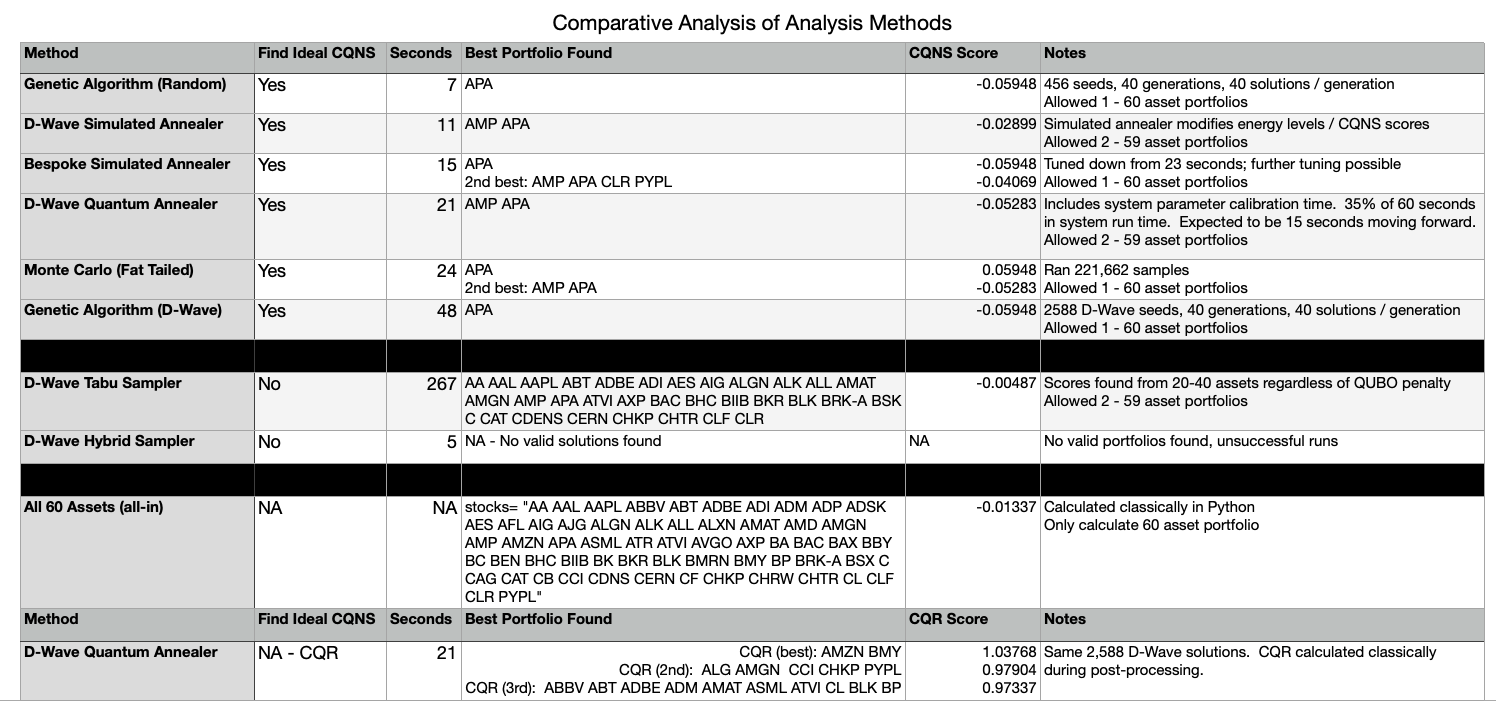}
	\caption{Full Table of Comparative Methods}
	\label{fig: comparison table full}
\end{figure}

\begin{figure}[!ht]
	\centering
	\includegraphics[width=1\textwidth]{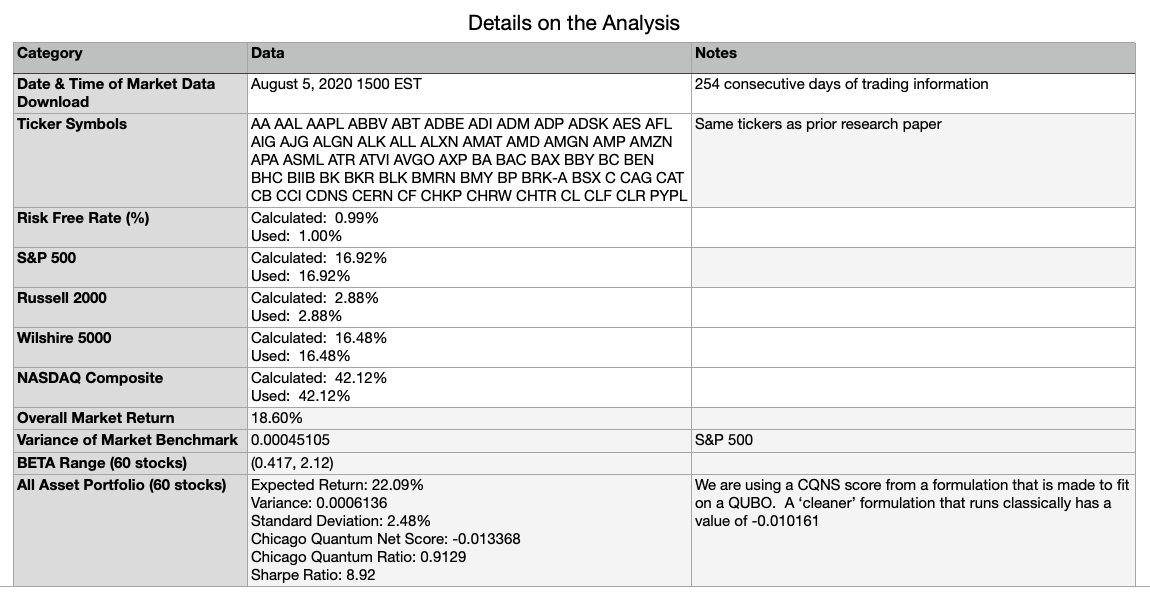}
	\caption{Details of Analysis}
	\label{fig: ticker deatils}
\end{figure}

Below are two charts \ref{fig: comparison table full}, \ref{fig: ticker deatils} with more complete details of our analyses, stock info, date and time, and portfolios selected by various methods.  We provide these details for the sake of repeatability by interested readers.

\end{document}